\documentclass[12pt]{article}
\usepackage{bm}
\usepackage{eucal, amssymb,amsmath,graphicx, amsfonts, latexsym}
\usepackage{natbib}
\begin{document} \parskip=5pt plus1pt minus1pt \parindent=0pt

\newcommand{\E}{{\rm E}}
\newcommand{\Var}{{\rm Var}}
\newcommand{\re}{{\rm e}}
\newcommand{\rg}{{\rm g}}
\newcommand{\rh}{{\rm h}}
\newcommand{\ru}{{\rm u}}
\newcommand{\rv}{{\rm v}}
\newcommand{\bn}{{\mbox{\boldmath $n$}}}
\newcommand{\ba}{{\mbox{\boldmath $a$}}}
\newcommand{\bA}{{\mbox{\boldmath $A$}}}
\newcommand{\btheta}{\bm{\theta}}

\title{Statistical inference for stochastic epidemic models with three levels of mixing}
\author{Tom Britton\thanks{Department of Mathematics, Stockholm
University, SE-106 91 Stockholm, Sweden.}, Theodore Kypraios\thanks{School
of Mathematical Sciences, University of Nottingham, Nottingham NG7 2RD, UK} and Philip
O'Neill\thanks{School of Mathematical Sciences, University of Nottingham,
Nottingham NG7 2RD, UK.}}
\date{}
\maketitle

\begin{abstract}
\noindent A stochastic epidemic model is defined in
which each individual belongs to a household, a secondary grouping (typically school or workplace)
and also the community as a whole. Moreover, infectious contacts take place in
these three settings according to potentially different rates.
For this model we consider how different kinds of data
can be used to estimate the infection rate parameters with a view to understanding
what can and cannot be inferred, and with what precision. Among other things we
find that temporal data can be of considerable inferential benefit compared to final
size data, that the degree of heterogeneity in the data can have a considerable
effect on inference for non-household transmission, and that inferences can be
materially different from those obtained from a model with two levels of mixing.
\end{abstract}


\vskip20pt
\section{Introduction}

Classical early work in mathematical epidemic modelling usually
assumed a homogeneously mixing community of individuals, each having
the same susceptibility to disease and the same ability to transmit disease
(see e.g.\ Kermack and McKendrick, 1927, and Bailey, 1975).
Such assumptions rarely reflect reality, and during the last thirty
years or so, considerable effort has been focused towards modelling
different heterogeneities in the community in question and their
effects on disease propagation (e.g.\ Anderson and May, 1991,
Diekmann and Heesterbeek, 2000, Keeling and Rohani, 2007, and
references therein). Heterogeneities can, broadly speaking, be
separated in two different types: individual heterogeneities (e.g.\
susceptibility and infectivity) and social heterogeneities caused by
the contact structures in the community (e.g.\ children attending
schools). For stochastic models, which are the focus of the present
paper, the first type of heterogeneity is usually represented by
multitype epidemics (e.g. Anderson and Britton, 2000, Chapter 6).
Heterogeneities caused by social structures have been incorporated
in relatively simple models by assuming mixing at different levels, such as community
mixing and mixing within households (Ball \emph{et al.}, 1997),
but also using random network models (e.g.\ Andersson, 1999, Britton
and O'Neill, 2002, and Newman, 2003).

More recently, there has been considerable interest in understanding
disease spread by using models that attempt to capture
the essential features of real-life human populations
(e.g.\ Ferguson \emph{et al.}, 2005, Longini \emph{et al.}, 2005, Halloran
\emph{et al.}, 2008). Mathematical models used to address these issues are
typically quite complex, incorporating information of various sorts
that may affect disease propagation in a given community. Examples include population
age structure, school sizes and household sizes, geographic
locations of villages and hospitals, information about travel (local
and national), disease characteristics such as latent and
infectious periods, possible interventions, and much more. Models
of this kind are then studied via intensive computer simulation to
identify key features of disease propagation and mitigation.

In order to make such complex models as realistic as possible, it is
vitally important to assign plausible values to the many parameters in the model. Typically,
some parameters are well-informed by data from existing studies,
while others are not. An important example of the latter are
parameters that govern mixing at an intermediate level, meaning
neither within-household or population-at-large mixing. More
precisely, not enough is known about the relative importance of
disease transmission at schools and workplaces (and similar) as
compared with transmission within households and more random type
transmission at community level. However, such information is vital
to public health policy, since control measures such as school
closures or restrictions on large public gatherings are aimed
precisely at reducing such intermediate transmission (see e.g. Cauchemez
{\em et al.}, 2009). A recent study in this area is Cauchemez
{\em et al.}\ (2008) in which model-based methods
are used to estimate the relative importance of transmission of influenza within
schools from longitudinal endemic data by comparing the number of reported cases during school term with the
number of reported cases during school holidays. The authors found
that approximately 25\% of all transmission among children came via
schools, which underlines the potential importance of intermediate-level
mixing in disease transmission.

This paper has two main aims. The first is to establish statistical
estimation procedures for models featuring an intermediate level of
mixing, given data on a single outbreak (note this is distinct from
the longitudinal data considered by Cauchemez \emph{ et al.}, 2008).
The second aim is to use these procedures to assess in broad
terms what can and cannot be estimated from outbreak data, using
simulated data. A stochastic epidemic model
incorporating three levels of mixing (households, intermediate and
community level) is defined. Inference procedures are then derived
for model parameters assuming two possible kinds of data, namely
final size data, consisting simply of case numbers, or complete
observation of the epidemic process through time. These two kinds of data
represent extremes corresponding to minimal or maximal observation,
and we consider them in order to ascertain what can and
cannot be estimated from actual observational data.

We consider simulated data from two 3-level mixing scenarios: first where a
community of households is separated into villages, and secondly
where households consists of adults and children, the former going
to workplaces and the latter to schools. In the first example all
members of a household belong to the same group structure (village),
whereas as in the second example members of a household belong to
different group structures (schools/workplaces). For both scenarios
the simulated data is then used to assess what can be inferred in terms of
the model parameters.

The paper is structured as follows. In Section 2 we define the
epidemic model with three levels of mixing. In Section 3 the
likelihood is derived for the case of complete observation of the
epidemic, and inference procedures for final size data are also discussed.
Sections 4 and 5 are devoted to the two specific examples of community
structures mentioned above, and we finish with conclusions and discussion in Section 6.

\section{The model}
\subsection{Definition and notation}
Consider a community consisting of $N$ individuals. Each individual belongs to
one household and also to exactly one group of a specific type, such as a school or
workplace. In the sequel we use the terminology {\em household}, {\em group} and
{\em community} to refer to the three populations to which an individual belongs.
Note that households may consist of different numbers of individuals,
and the same applies to groups.

Suppose that the households are labelled $1, \ldots ,n$, and the
different groups by $0, 1, \ldots, J$. Here group 0 is a dummy group:
individuals that do not belong to a group are said to belong to group 0.
An individual may thus be thought of as being type
$(i,j)$, meaning that they are in household $i$, and group $j$. As
described below, these types create potential differences between
individuals in terms of their mixing behaviour. However, individuals
are otherwise assumed to be similar, meaning that all are equally
susceptible to the disease and equally able to infect others. Thus
the population is homogeneous apart from the mixing behaviour of
individuals. Finally, note that the only potential difference
between individuals in the same household is their group membership.

The model is of SEIR (Susceptible-Exposed-Infective-Removed) type
(Diekmann and Heesterbeek, 2000), meaning that at any point in time,
each individual in the population is either susceptible, exposed,
infective, or removed. Susceptible individuals have the potential to
contract the disease. Exposed (or latent) individuals have been
infected but are not yet capable of transmitting the disease to
others. Infectives can transmit the disease to others, while removed
(or recovered) individuals are no longer infectious, and moreover
immune to further infections.

An individual who becomes infected first enters the exposed (or latent)
period whose duration is distributed according to some specified
non-negative random variable $T^{(E)}$. Following this, the
individual becomes infective, remaining so for a period of time
distributed according to some specified random variable $T^{(I)}$
with mean $E(T^{(I)}) = \mu$. The exposed and infectious periods of
a single individual and of different individuals are all assumed to
be mutually independent. We denote by $f_E$ and $f_I$, respectively,
the probability density (or mass, as appropriate) functions of the
exposed and infectious periods, and denote by $\bar F_E$ and $\bar
F_I$ the corresponding survivor functions ($\bar
F(t)=1-F(t)=\int_t^\infty f(s)ds, \; t \geq 0$).

During its infectious period an individual may make three types of
contact according to various mutually independent Poisson
processes of different rates as described shortly. All such contacts that take place with susceptible
individuals result in the immediate infection of that individual, so
that the contacted individual enters the exposed period. First,
an infectious individual has contacts with each household member independently at
times given by the points of a homogeneous Poisson process of rate
$\lambda_{\rm H}$. Second, if the infectious individual belongs to group $j$, then
the individual has contacts with each individual in the same group
according to a Poisson process of rate $\lambda_{\rm G}^{(j)}/n_j$, where $n_j$ denotes the number
of individuals in the group. We set $\lambda_{\rm G}^{(0)} = 0$ because
group 0 is a dummy group. Note that the contact rate in different
groups can vary. Finally, an infectious individual also has contacts
with each individual in the entire community according to a Poisson
process of rate $\lambda_{\rm C}/N$.

Initially, the population consists of one or a few exposed or
infective individuals, with all other individuals being susceptible
(initially immune individuals are simply be ignored). The epidemic continues
until there are no exposed or infective individuals present in the
community. Each individual is then either still susceptible, or else
they have been infected and have recovered.

Note that the parameters $\lambda_{\rm C}$ and $\lambda_{\rm G}^{(j)}$
represent the overall rates that an individual (in group $j$) has community and
group contacts, respectively. Conversely, the overall rate of household
contacts is $(h-1)\lambda_{\rm H}$ in a household of size $h$, following
the convention of Ball \emph{et al.}\ (1997).

\subsection {Threshold behaviour}
Stochastic epidemic models typically exhibit threshold behaviour,
meaning that in a large population, epidemics either die out
quickly, or else may infect a non-negligible fraction of the
population with positive probability (e.g.\ Andersson and Britton,
2000). Moreover, this dichotomy is characterised via some threshold
parameter $R_*$, itself a function of the model parameters, with
$R_*=1$ the boundary between the two behaviour regimes: when $R_*\le
1$ the epidemic will die out quickly with certainty whereas when
$R_*>1$ it can either die out quickly or else a non-negligible
fraction of the population is infected. The quantity $R_*$, which is
a so-called reproduction number, is of key practical importance because
control strategies typically aim to reduce its value to below the
critical value of 1.

Threshold behaviour for the above model, when
the population size $N$ is large, can be derived in a manner similar to
that for the multitype two-level mixing model defined in Ball and Lyne
(2001). We now give a brief outline of the argument; specifics are described
later for the particular examples considered in this paper.

Suppose that the population size, $N$, is assumed to become large in
such a way that household sizes remain unchanged, but group sizes
become large. Note that this can happen in various ways; for
instance, the number of groups may be fixed, or may also increase.
However, the number of different types of group, i.e.\ the number of
distinct values of $\lambda_{\rm G}^{(j)}$, is assumed to remain
fixed. The key requirement is that, as $N \rightarrow \infty$ and
for any fixed time $t$, the probability that a given household
receives more than one infectious contact from outside the household
before $t$ tends to zero. This means that the process of infections
between households can be regarded as a branching process in which an individual corresponds
to a household and birth corresponds to infection, the point here
being that each non-household infection that occurs will be with an
individual in a previously uninfected household.

The threshold parameter for this branching process constitutes a
threshold parameter for the epidemic model. In general, the
branching process will be multitype, where the different types
reflect the group of the first individual in a household to
become infected as well as the group type of all household members
(cf. Ball and Lyne, 2001). By defining $m_{ij}$ as the mean number
of type $j$ offspring from a type $i$ individual in the branching
process (where `type' refers to the household structure just described), the threshold parameter equals the maximal eigenvalue of
the matrix $M = (m_{ij})$.

In Example 1 of Section 4 below, we consider the simple case of
equal-sized households and all groups having the same transmission
rate $\lambda_G$. In this setting the threshold
parameter is simply
\[
R_*  =(\lambda_C + \lambda_G) E(T_A),
\]
where $T_A$ is the final severity of a single isolated household
containing one initial infective (i.e.\ the sum of the infectious
periods of those ever infected). It can be shown (Ball \emph{ et al.},
1997) that $E(T_A) = \mu E(T)$, where $T$ denotes the final number
of individuals in the household ever infected, including the initial
infective, and recalling that $\mu =E(T^{(I)})$. Note also that the threshold parameter is independent of
the exposed period. In Example 2 in Section 5 we derive $R_*$ for a
more complicated model.

\section{Statistical inference}

We now turn our attention to statistical inference, specifically
considering two kinds of dataset. The first consists of complete
temporal information, i.e.\ knowledge of the state of every individual
in the population throughout the disease outbreak. The second type of data treated consists
only of observation at the start and end of the epidemic, i.e.\ knowledge of
who was initially susceptible and which of these individuals were infected
during the outbreak. Our motivation for focussing
on these two types of data is that they represent extreme scenarios.
In practice, actual outbreak data is likely to be somewhere in between,
for instance observing just removals through time, or weekly aggregates
of case numbers. By considering the two data types described above, we
can gain insight into what can and cannot be estimated even in extreme
cases, and also evaluate the benefit of more detailed data collection.

Finally, it is also assumed that the social structures are known, i.e.\ that we
know which household and secondary grouping each individual belongs to.
In practice this assumption is not unreasonable, but even so our methods
could be extended to take account of missing data on social structures.

\subsection{Likelihood based inference of complete data}

We now derive both the likelihood for the complete data and
maximum-likelihood estimates for the parameters using counting
process theory (e.g.\ Andersen \emph{et al.}, 1993). For $t \geq 0$
let $I(t)$, $I^H_i(t)$ and $I^G_j(t)$ denote the number of infective
individuals in the community, in household $i$ and in group $j$,
respectively, at time $t$. Let $S(t)$, $S^H_i(t)$ and $S^G_{j}(t)$
denote the corresponding susceptible numbers at time $t$, and
further define $S^{H,G}_{i,j}(t)$ as the number who are in both
household $i$ and group $j$, and susceptible at time $t$. Thus
$\sum_{j = 0}^{J} S^{H,G}_{i,j}(t) = S^H_i(t)$. We use the notation
$t-$ to denote the left limit, e.g.\ $I(t-) = \lim_{s \uparrow t}
I(s)$. Let $t_{i,j,r}$ denote the time of the $r$th infection among
individuals in household $i$ belonging to group $j$, with the
convention that $t_{i,j,r}=\infty$ when no such infection occurs.
The length of the corresponding infected individual's exposed and infectious
periods are denoted $T^{(E)}_{i,j,r}$ and $T^{(I)}_{i,j,r}$,
respectively. Assuming that the period of observation is $[0,t]$,
the likelihood is given by
\begin{eqnarray}
L(\lambda_{\rm H}, \lambda_{\rm G}, \lambda_{\rm C};t)&=&\prod_{\{i,j,r:\
t_{i,j,r}< t \}}\left[S^{H,G}_{i,j}(t_{i,j,r}-)\left(
\lambda_{\rm H}I^H_{i}(t_{i,j,r}-) + \lambda_{\rm G}^{(j)}
\frac{I^G_{j}(t_{i,j,r}-)}{n_{j}} + \lambda_{\rm C}
\frac{I(t_{i,j,r}-)}{N}\right)\right]\nonumber\\
&&\times \exp{\left[-\int_0^t \left(\sum_i\lambda_{\rm H}S^H_i(s)I^H_i(s)
+\sum_{j}\lambda_{\rm G}^{(j)}\frac{S^G_{j}(s)I^G_{j}(s)}{n_{j}}
+\lambda_{\rm C}\frac{S(s)I(s)}{N}  \right) \, ds\right]}\nonumber
\\&&\times \prod_{\{i,j,r \}} \chi(i,r,j,t),\label{lik}
\end{eqnarray}
where
\[
\chi(i,j,r,t) =
\left\{
\begin{array}{ll}
1 & \mbox{if $t < t_{i,j,r}$,}\\
\bar F_E (t - t_{i,j,r}) & \mbox{if $t_{i,j,r} \leq t < t_{i,j,r} +
  T^{(E)}_{i,j,r}$,}\\
f_E(T^{(E)}_{i,j,r})\bar F_I(t - t_{i,j,r} - T^{(E)}_{i,j,r}) &
\mbox{if $t_{i,j,r} + T^{(E)}_{i,j,r} \leq t < t_{i,j,r} +
  T^{(E)}_{i,j,r} + T^{(I)}_{i,j,r}$,}\\
f_E(T^{(E)}_{i,j,r}) f_I (T^{(I)}_{i,j,r}) & \mbox{otherwise.}
\end{array}
\right.
\]

From (\ref{lik}) it is possible to derive maximum likelihood (ML)
estimates for the contact rates $\{ \lambda_{\rm H},\ \lambda_{\rm
G}^{(j)};\ j=1,\ldots, J,\ \lambda_{\rm C}\}$, as described below. It is
also straightforward to obtain estimates of the parameters of the
exposed and infectious period distributions, since from (\ref{lik})
the complete data provide (potentially censored) independent
and identically distributed observations from the two
distributions. The last product (containing the factors
$\chi(i,j,r,t)$) carries all information about the
latent and infectious periods. Thus, when focus lies in making inference
about transmission parameters or when the infectious and latent
distributions are known, this product can be neglected.

To make inference of the contact parameters we use the
log-likelihood $\ell =\ln(L)$ and differentiate it with respect to
each parameter separately, yielding
\begin{eqnarray*}
\frac{\partial \ell}{\partial \lambda_{\rm H}}&=&
\sum_{ \{ i,j,r : \; t_{i,j,r}< t \} } \frac{I^H_{i}(t_{i,j,r}-)}{
\lambda_{\rm H}I^H_{i}(t_{i,j,r}-) + \lambda_{\rm
G}^{(j)}I^G_{j}(t_{i,j,r}-)/n_j + \lambda_{\rm
C}I(t_{i,j,r}-)/N} \\&&\hskip2cm - \int_0^t
\sum_i S^H_i(s) I^H_i(s)\,ds ,
\\
\frac{\partial \ell}{\partial \lambda_{\rm G}^{(j)}} &=&
\sum_{ \{i,r: \; t_{i,j,r}<t \}}\frac{I^G_{j}(t_{i,j,r}-)/n_{j}}{
\lambda_{\rm H}I^H_{i}(t_{i,j,r}-) + \lambda_{\rm
G}^{(j)}I^G_{j}(t_{i,j,r}-)/n_{j} + \lambda_{\rm
C}I(t_{i,j,r}-)/N} \\&&\hskip2cm - \int_0^t
\sum_k (S^G_{j}(s) I^G_{j}(s)/n_{j})\,ds ,
\\
\frac{\partial \ell}{\partial \lambda_{\rm C}}&=&
\sum_{\{i,j,r : \; t_{i,j,r}< t \}}\frac{I(t_{i,j,r}-)/N}{
\lambda_{\rm H}I^H_{i}(t_{i,j,r}-) + \lambda_{\rm
G}^{(j)}I^G_{j}(t_{i,j,r}-)/n_{j} + \lambda_{\rm
C}I(t_{i,j,r}-)/N} \\&&\hskip2cm - \int_0^t (S(s)I(s)/N) \, ds .
\end{eqnarray*}
Note that the first summation in the expression for $\partial
\ell/\partial \lambda_{\rm G}^{(j)}$ does not extend over $j$. To
obtain the ML-estimates these equations are set equal to 0 and are
then solved in terms of the parameters. There are $J+2$
equations and equally many unknowns (parameters). Quite often in
large communities several groups are of the same type, such as
schools or villages, thus having the same $\lambda_G$. Then the
corresponding partial derivatives should be summed up thus reducing
the number of equations to two plus the number of different types of
groupings. If this number is small, maximum likelihood estimates
are easy to obtain numerically.

Finally, Bayesian inference is straightforward given the likelihood
defined at (\ref{lik}), for instance by using a Metropolis-Hastings
algorithm to obtain approximate samples from the joint posterior
distribution of the unknown infection rate parameters. Details of
such inference will be given in the specific examples below.

\subsection{Inference for final size data}
In contrast to complete temporal data, we also consider final size
data. Such data consist simply of case numbers at the end of
the epidemic outbreak, so that there is no explicit temporal information.
For our three-level mixing model, statistical inference based on
final size data is generally far more challenging than for complete
temporal data, the reason being that it is often impractical to evaluate
(both analytically and numerically) the required likelihood function.
Although methods exist to overcome this problem (notably data augmentation
MCMC methods, see e.g.\ Demiris and O'Neill, 2005 and O'Neill, 2009), here we shall adopt a
simpler approach by using an approximation in which households behave
independently of one another. Such approximations are common in the
inference literature for two-level mixing models (as discussed in
Demiris and O'Neill, 2005), and are frequently reasonable in practice,
especially in large populations. The details are given below.

\subsection{Latent and infectious periods}
As mentioned above, with complete data it is a simple matter to perform
statistical inference for parameters governing the latent and infectious
period distributions. Conversely, given final size data it is impossible
to estimate latent period parameters, since the final size distribution is
itself invariant to the choice of latent period (see e.g.\ Ball \emph{ et al.},
1997). Moreover, the final size distribution is not greatly affected by
the choice of infectious period distribution other than through its mean,
and so most realistic choices of infectious period
distribution result do not materially affect inferences for infection rate
parameters (see e.g.\ O'Neill \emph{ et al.}, 2000). In view of these facts,
in the numerical illustrations in the sequel we shall assume that both
latent and infectious periods are simply fixed, i.e.\ non-random. In particular,
this means that estimation of infection rate parameters cannot be confounded
by uncertainty in the estimation of latent or infectious period distribution parameters.
However, some derivations of quantities of interest (pseudolikelihoods etc.) will
be given in the general case, i.e.\ arbitrary distributions for latent and
infectious periods.

\section{Example 1: Households of size 2 in villages}
We start with a fairly simple example that still allows us to explore questions of
interest, such as what can be estimated, and with what precision.

\subsection{Model and threshold parameter}
Consider a population of $n$ households of size 2, all individuals
being of the same type. There are $m$ villages, all of the same
size, and we assume further that the mixing rate within villages is
the same for different villages, so $\lambda_G^{(j)}=\lambda_G$ for
all $j$. It follows that $N=2n$ and that
the villages consist of $N/m$ individuals each.

As mentioned in Section 2.2, the threshold parameter for this model
is given by $R_* = (\lambda_{\rm C}+ \lambda_{\rm G})\mu E(T)$,
where $T$ is the total number of infected individuals in a household
in which one individual becomes infected. It follows that
\[
R_* = (\lambda_{\rm C}+ \lambda_{\rm G})\mu (1+p_{\rm H})
\]
where $p_{\rm H}=(1-\E(e^{-\lambda_{\rm H}T^{(I)} } ))$ is the probability
that a single infected individual infects the other individual in its household
via household transmission.

\subsection{Approximate statistical inference from final size data}

Assume now that an outbreak in the community has occurred and that
the final size has been observed. Since all households contain two
individuals who have the same group membership, the data
can be summarized as $\bn = \{ \bn_j = (n_0^{(j)},n_1^{(j)},
n_2^{(j)}) : j=1,\ldots ,m\}$, where $n_k^{(j)}$ denotes the number
of households in village $j$ in which $k$ individuals were infected during the
epidemic outbreak.

Performing inference for $\lambda_{\rm C}$, $\lambda_{\rm G}$ and
$\lambda_{\rm H}$ (or equivalently $p_{\rm H}$) is possible in the
Bayesian context by adopting the random graph imputation approach
described in Demiris and O'Neill (2005). However, here we focus on a
much quicker and simpler approach that nevertheless enables us to
address the question of what can be estimated.

For $j=1, \ldots, m$, let $Z_j$ denote the number of ultimately infected individuals in
village $j$, so that $Z_j = \sum_{k=0}^{2} k n_k^{(j)}$, and define
$\bar{Z_j} = Z_j / (N/m)$ as the corresponding proportion of
infected individuals in the village. Similarly,
define the proportion of the entire community that is ultimately
infected by $\bar{Z} = \sum_{j=1}^{m} Z_j / N$. By neglecting the
dependence between households we can obtain a pseudolikelihood for
the data as follows. First note that the probability that an
individual avoids infection from a single infective via community infection
is $E[e^{- \lambda_C T^{(I)} /N}]$. As described above, we assume for simplicity that
$T^{(I)} \equiv \mu$ is nonrandom.
It then follows that the probability that a susceptible individual avoids community infection
from $k$ infectives is $e^{- \lambda_{\rm C} \mu k/N}$. By
neglecting the dependencies inherent in the model, it follows that
the probability that $j$ susceptibles avoid community infection from
$N \bar{Z}$ infectives is $e^{- j\lambda_{\rm C} \mu N \bar{Z} / N}
= \pi_{\rm C}^{ j\bar{Z} }$, say, where $\pi_{\rm C} := e^{-
\lambda_{\rm C} \mu}$.  Similar arguments hold for group infections,
and we define $\pi_{\rm G} = e^{- \lambda_G \mu}$. The
pseudolikelihood for the data using the new parameters is therefore
\begin{equation}
L( \{ \bn_j \}_{j=1}^m ; \ p_{\rm H}, \pi_{\rm C}, \pi_{\rm
G} )= \prod_{j=1}^m[\pi_j^2]^{n_0^{(j)}}[2\pi_j(1-\pi_j)(1-p_{\rm
H})]^{n_1^{(j)}}[2\pi_j(1-\pi_j)p_{\rm H}+(1-\pi_j)^2]^{n_2^{(j)}},\label{pseudo1}
\end{equation}
where for $j=1, \ldots, m$, $\pi_j=e^{-(\lambda_{\rm C}\mu\bar
Z+\lambda_{\rm G}\mu\bar Z_j)}=\pi_{\rm C}^{\bar Z}\pi_{\rm G}^{\bar
Z_j}$ is the approximate probability that an individual in group $j$
avoids infection from both group and community. Finally,
note that $R_* = - (\log \pi_{\rm C} + \log \pi_{\rm G})(1 + p_H)$
when written in terms of the new parameters.

\subsection{Bayesian inference}

Bayesian analyses of both the complete and final size data were performed
using MCMC methods as described below. For the final size data, the
three model parameters $p_{\rm H}$, $\pi_{\rm G}$ and $\pi_{\rm C}$
were assigned independent $U(0,1)$ prior distributions. For the complete data,
it is more natural to work with the original model parameters and the
likelihood at (\ref{lik}). The prior distributions on the original parameters
were set to be independent exponential with mean 1, since these are equivalent to the
$U(0,1)$ prior distributions on the new parameters.
For both complete and final size data,
Bayesian inference is based on the joint posterior density of the model
parameters given the data. This density is defined, up to proportionality,
by the product of the prior density and the appropriate likelihood, namely
(\ref{lik}) for the complete data and the pseudolikelihood
(\ref{pseudo1}) for the final size data.

For both complete and final size data, analysis of the posterior density of
interest was performed using a Metropolis-Hastings algorithm in which each
of the three model parameters was updated separately. For final size data,
the proposal distribution was $U(0,1)$, which gave adequate mixing in practice.
For the complete data, a random-walk algorithm was used in which the proposal
distribution was Gaussian, centered on the current parameter value.

\subsection{Results}
We present results using two different simulated datasets, described below.
In both cases we assume that there are $m=4$ identical villages, each
consisting of 500 households of size two, that the infectious period is
$T_I \equiv \mu = 1$ and that the epidemic outbreak is initiated by one individual (in village 1) being infected from outside.

{\bf Dataset 1.1} The first data set was simulated from the model with $\lambda_{\rm H} = 0.3$,
$\lambda_{\rm G} = 1.4$ and $\lambda_{\rm C} = 0.001$, which yields true parameter
values $p_{\rm H} = 0.259$, $\pi_{\rm G} = 0.247$ and $\pi_{\rm C} =
0.999$. It follows that the threshold parameter is $R_*=(\lambda_{\rm C} +
\lambda_{\rm G} )\mu(1+p_{\rm H})=1.763$. The parameter values were chosen to reflect a community in which mixing is quite high within villages and much less between villages.

The simulated outbreak (itself a fairly typical outbreak for the chosen
parameter values) resulted in the following final outcome data set:
\[
\bn_1 = (70,157,273), \; \; \bn_2 = (65,178,257), \; \; \bn_3 =
(500,0,0), \; \; \bn_4 = (500,0,0).
\]
Thus $n_0^{(1)} = 70$, $n_1^{(1)} = 157$, $n_2^{(1)} = 273$, for
example.

\begin{table}
  \centering
  \caption{Posterior density summaries and ML estimates, Dataset 1.1}\label{ex1.1}
\vskip15pt
\begin{tabular}{l|c|c c c c |c c c c}
&True& \multicolumn{4}{|c}{Complete data} & \multicolumn{4}{|c}{Final size data}\\
&value& Mean & S. Dev. & Median & MLE & Mean & S. Dev. & Median & MLE\\
\hline
$p_H$ &0.259& 0.264 & 0.017 & 0.264 & 0.263 & 0.277 & 0.037 & 0.278 & 0.279\\
$\pi_{\rm G}$ &0.247& 0.232 & 0.010 & 0.233 & 0.232 & 0.238 & 0.014 & 0.238 & 0.238\\
$\pi_{\rm C}$ &0.999& 0.998 & 0.002 & 0.998 & 0.999 & 0.999 & 0.001 & 0.999 & 1.000\\
$R_*$ &1.763& 1.847 & 0.058 & 1.846 & 1.847 & 1.836 & 0.062 & 1.835 & 1.836
\end{tabular}
\end{table}

\begin{table}
\centering
\caption{Posterior correlations, Dataset 1.1}\label{ex1.1cor}
\vskip15pt
\begin{tabular}{l|c|c}
& Complete data & Final size data\\
\hline
$\rho(p_H, \pi_{\rm G})$ & 0.069 & 0.57\\
$\rho(p_H, \pi_{\rm C})$ & 0.00037 & 0.0014\\
$\rho(\pi_{\rm G}, \pi_{\rm C})$ & -0.0018 & -0.0080\\
\end{tabular}

\end{table}

{\bf Dataset 1.2} The second data set was simulated from the model with $\lambda_{\rm H} = 0.3$,
$\lambda_{\rm G} = \lambda_{\rm C} = 0.6$ giving true parameter
values $p_{\rm H} = 0.259$, $\pi_{\rm G} = \pi_{\rm C} = 0.549$ and
threshold parameter $R_* = 1.511$. The difference as compared to the parameter values of the first dataset is that now the mixing within villages has decreased and community mixing has increased.

The simulated dataset, again typical, was
\[
\bn_1 = (137,180,183), \; \; \bn_2 = (114,182,204), \; \; \bn_3 =
(128,177,195), \; \; \bn_4 = (126,188,186).
\]
In contrast to dataset 1.1, here each village undergoes a similar outbreak.

\begin{table}
  \centering
  \caption{Posterior density summaries and ML estimates, Dataset 1.2}\label{ex1.2}
\vskip15pt
\begin{tabular}{l|c|c c c c |c c c c}
&True& \multicolumn{4}{|c}{Complete data} & \multicolumn{4}{|c}{Final size data}\\
&value& Mean & S. Dev. & Median & MLE & Mean & S. Dev. & Median & MLE\\
\hline
$p_H$ &0.259& 0.270 & 0.0012 & 0.270 & 0.269 & 0.272 & 0.021 & 0.272 & 0.272\\
$\pi_{\rm G}$ &0.549& 0.529 & 0.046 & 0.529 & 0.525 & 0.498 & 0.173 & 0.451 & 0.260\\
$\pi_{\rm C}$ &0.549& 0.563 & 0.049 & 0.561 & 0.565 & 0.658 & 0.195 & 0.656 & 1.000\\
$R_*$ &1.511& 1.545 & 0.038 & 1.545 & 1.542 & 1.550 & 0.040 & 1.549 & 1.713
\end{tabular}
\end{table}

\begin{table}
\centering
\caption{Posterior correlations, Dataset 1.2}\label{ex1.2cor}
\vskip15pt
\begin{tabular}{l|c|c}
& Complete data & Final size data\\
\hline
$\rho(p_H, \pi_{\rm G})$ & 0.012 & 0.025\\
$\rho(p_H, \pi_{\rm C})$ & 0.0032 & 0.012\\
$\rho(\pi_{\rm G}, \pi_{\rm C})$ & -0.94 & -0.95\\
\end{tabular}

\end{table}

{\em General remarks on estimation} Tables \ref{ex1.1} - \ref{ex1.2cor}
contain the results of the MCMC analyses
and maximum likelihood estimates for the two datasets.
The true parameter values are given for reference, although since
inference is based on just one simulation it follows that point
estimates are not expected to be identical to true values.

The estimates in Table \ref{ex1.1}
illustrate that final size data alone can be sufficient to yield
reasonable estimates of the three model parameters. In addition, the
approximate approach to inference using a pseudolikelihood for the final
size data appears to be effective in this case. As would be expected,
complete data based estimates are usually more precise than final size data
based estimates in the sense that the former have lower posterior
standard deviations than the latter. The posterior standard deviations of the
model parameters are reduced by approximately 50\% for dataset 1.1 and
even more for dataset 1.2. Finally, both point and uncertainty estimation
for the threshold parameter $R_*$ is similar for both kinds of data, illustrating
that inference about $R_*$ can be effectively performed with final size data alone.
Similar findings for standard SIR models are described in Clancy and O'Neill (2008).

{\em Data heterogeneity affects estimation} A key finding is that
the extent to which the three model parameters can be individually
estimated is dependent upon the between-village diversity in the data.
In dataset 1.1, two villages undergo outbreaks whilst the other two
remain completely unaffected. Conversely, in dataset 1.2 all four
villages experience similar outbreaks. Intuitively, this similarity
makes it harder to differentiate the roles of $\pi_{\rm G}$ and
$\pi_{\rm C}$, since either group or community infection, or both,
could be driving the epidemic. This is reflected in the fact
that the posterior correlation $\rho(\pi_{\rm G}, \pi_{\rm C})$ for
dataset 1.2 is close to -1 for both final size and complete data based estimation.
Conversely, in dataset 1.1 the corresponding posterior correlations
are far lower, illustrating that here it is feasible to distinguish
between group and community infection.

{\em Maximum likelihood vs. Bayesian estimation} It is interesting
to note that maximum likelihood estimation performs poorly for dataset
1.2 when considering final size data, where the estimates for
$\pi_{\rm C}$ and $\pi_{\rm G}$ are nothing like either the true values
or the Bayesian posterior averages. It seems likely that this occurs due
to the shape of the likelihood surface, itself a consequence of the correlation
between $\pi_{\rm G}$ and $\pi_{\rm C}$ discussed above. These findings also
highlight the need for caution in interpreting ML estimates.

{\em Are three levels of mixing really necessary?} It is natural to compare
the differences in inference based on three-level mixing models with
that based on a two-level mixing model in which group-level mixing is ignored,
i.e.\ the within-group infection rate is set to zero. For dataset 1.1,
by setting $\pi_{\rm G} = 0$ the final size analysis yields
$E[p_H | \bn] = 0.549$, $E[\pi_{\rm C} | \bn] = 0.549$  and
$E[R_* | \bn] = 1.26$. Two points should be noted. First, even though
we still have the same household-level data, the estimate of the
within-household transmission probability $p_H$ is markedly different
under the assumption of a two-level mixing model. Second, one might at
least hope that inference for the threshold parameter $R_*$ was relatively
robust to the choice of model, but again the choice of model has, in
this case, a considerable impact.

\section{Example 2: Households and school/workplaces}

In our second example we consider a population divided into households
of equal size, where each household contains individuals belonging to one of
two kinds of groups, which represent schools and workplaces.

\subsection{Model and threshold parameter}
The mixing structure of the model is defined as follows.
The population is partitioned into $n=500$ households, all of size 4.
There are two secondary group structures, type 1 being schools with
contact rate $\lambda_{\rm G}^{(1)}$ and type 2 being workplaces
with contact rates $\lambda_{\rm G}^{(2)}$. In each household two
individuals go to school and two go to a workplace. For convenience we
refer to the former two individuals as children, and the latter two as
adults, one male and one female. There are
$m_1=10$ schools, each of size 100 (= $2n/m_1$), and $m_2=40$
workplaces each of size 25 (= $2n/m_2$).

We assumed the following allocation of children to schools and adults
to workplaces. Suppose that the households,
schools and workplaces are numbered (e.g.\ $1, 2, \ldots, 500$ for
households). Children are allocated to schools such that school 1
is populated by all children from households 1-50, school 2 by all children from households
51-100, etc. (so siblings are in
the same school). The two adults in each household are allocated to different
workplaces: in each of households 1-25 the two adults go to workplaces 1 and 21 respectively;
in each of households 26-50 the two adults go to workplaces 2 and 22 respectively, etc.
Obviously, numerous other allocations (both specified and at-random) are
possible but the exact choice of allocation seems unlikely to have a
material impact on model parameter estimation.

We now derive the threshold parameter $R_*$ for this model by considering
the approximating branching process in which an individual corresponds to
a household in the epidemic model.
Consider first the epidemic within a household, meaning that group
and community transmission is ignored. Recall that within-household
transmission is defined such that there is no difference between adults
and children. The final size distribution in such a household,
given the number of intially infective individuals, can be derived
from standard recursive formulae (e.g.\ Andersson \& Britton, 2000, p 16).
For $s \geq 0$, $i \geq 1$ and $k=0, \ldots, s$, let
$p_s^{(i)}(k)$ denote the probability that exactly $k$ of $s$ initially
susceptible individuals become infected during the course of the household epidemic
initiated by $i$ infective individuals. Further, let $\mu_s^{(i)}=\sum_{k=0}^skp_s^{(i)}(k)$
denote the mean number of initially susceptible individuals
who ever become infected.

During the early stages of the epidemic, with high probability each household receives at most
one external infectious contact. The average total number of individuals
infected in a household in which one was externally infected is therefore
$1+\mu_3^{(1)}$. We define a household to be type 1 if the externally
infected individual is a child and type 2 if it is an adult.

We now derive the mean offspring matrix $M= ( m_{ij} )$, where $m_{ij}$ denotes the expected number of
type $j$ households that one type $i$ household infects, starting with $m_{11}$.
The average number of infected individuals in a type 1 household is
$1+\mu_3^{(1)}$. All these individuals have global contacts at rate
$\lambda_{\rm C}$ with randomly chosen individuals in the community,
so the community contact rate with children is $\lambda_{\rm
  C}/2$ because half of the community are children. The mean length of
the infectious period is $\mu = E(T^{(I)})$, so the expected number of
type 1 households infected through community contacts equals
$(1+\mu_3^{(1)})\mu\lambda_{\rm C}/2$. It is also possible to infect
other type 1 households through school contacts. The expected number of
children infected in the type 1 household equals $1+ \mu_3^{(1)}/3$, i.e.\ the
externally infected child, plus one third of the mean number of susceptibles
infected, since they comprise one child and two adults. Each infected child
infects on average $\mu \lambda_{\rm C}^{(1)}$ other children in their school. Under
the assumption that schools are large, two infected children in a household will
each infect different individuals at school with high probability. The average
number of new type 1 households the type 1 household infects through school
contacts hence equals $(1+ \mu_3^{(1)}/3)\mu \lambda_{\rm C}^{(1)}$.
The community and school terms together make up $m_{11}$. The remaining
$m_{ij}$ can be obtained similarly, yielding
\begin{equation}
M=
\left(
\begin{array}{cc}
(1+\mu_3^{(1)})\mu\frac{\lambda_{\rm C}}{2}+ (1+ \frac{\mu_3^{(1)}}{3})\mu
\lambda_{\rm G}^{(1)} & (1+\mu_3^{(1)})\mu\frac{\lambda_{\rm C}}{2}+
\frac{2\mu_3^{(1)}}{3}\mu \lambda_{\rm G}^{(2)}
\\
(1+\mu_3^{(1)})\mu\frac{\lambda_{\rm C}}{2}+
\frac{2\mu_3^{(1)}}{3}\mu \lambda_{\rm G}^{(1)} & (1+\mu_3^{(1)})\mu\frac{\lambda_{\rm C}}{2}+ (1+ \frac{\mu_3^{(1)}}{3})\mu
\lambda_{\rm G}^{(2)}
\end{array}
 \right).
\label{M-ex2}
\end{equation}
The reproduction number $R_*$ is the largest eigenvalue of $M$, so that
\begin{equation}
R_*=\frac{m_{11} + m_{22}}{2}+\sqrt{\frac{(m_{11} - m_{22})^2}{4} +m_{12}m_{21}},\label{R-ex2}
\end{equation}
where $m_{ij}$ is defined in Equation (\ref{M-ex2}).

\subsection{Approximate statistical inference from final size data}

We now derive a pseudolikelihood for the final size in a similar manner to
that described for Example 1 above.
Label the schools 1 to $m_1$ (=10) and the work
places 1 to $m_2$ (=40). Recall that there are $n=500$ households and
the community size is $N=2000$.
All schools have size $N_s=100$ and all workplaces have size $N_w=25$. Let
$n_s^{(i)}$ denote the number of children in school $i$ who ever became infected
and  $n_w^{(j)}$ the number of adults in workplace $j$ who ever became infected
($i=1,\dots ,m_1$, $j=1,\dots ,m_2$). Let $n_c$ denote the total number of
individuals in the community who ever become infected. All of the quantities defined
in this paragraph are known from the final size data.

Label the households 1 to $n$ (=500) and let $k_c(h)$, $k_f(h)$ and $k_m(h)$
respectively denote the school of the children, the work place of the female, and the
workplace of the male in household $h$. The respective probabilities
that these individuals avoid both group and community infection are, approximately,
\begin{align*}
\psi_c(h)&=\exp\left({-\mu \left(\lambda_{\rm C}\frac{n_C}{N}+\lambda_{\rm
      G}^{(1)}\frac{n_s^{(k_c(h))}}{N_s}\right) } \right) = \pi_{\rm C}^{n_C/N} (\pi_{\rm G}^{(1)})^{n_s^{(k_c(h))}/N_s},
\\
\psi_f(h)&= \exp\left({-\mu \left(\lambda_{\rm C}\frac{n_C}{N}+\lambda_{\rm
      G}^{(2)}\frac{n_w^{(k_f(h))}}{N_w}\right) } \right) = \pi_{\rm C}^{n_C/N} (\pi_{\rm G}^{(2)})^{n_s^{(k_f(h))}/N_s},
\\
\psi_m(h)&= \exp\left({-\mu \left(\lambda_{\rm C}\frac{n_C}{N}+\lambda_{\rm
      G}^{(2)}\frac{n_w^{(k_m(h))}}{N_w}\right) } \right) = \pi_{\rm C}^{n_C/N} (\pi_{\rm G}^{(2)})^{n_s^{(k_m(h))}/N_s},
\end{align*}
the first expression being the probability that a given child in the
household avoids external infection, etc, and where similarly to
the previous section, $\pi_{\rm C} = e^{- \lambda_{\rm C} \mu}$ and
$\pi_{\rm G}^{(j)} = e^{- \lambda_{\rm G}^{(j)} \mu}$ ($j=1,2$). We also
define $p_{\rm H}=1-e^{- \lambda_{\rm H} \mu }$ as the between-individual
transmission probability within a household, which is the same for all
households.

We now derive an expression for the probability
$p_h(i,j,k)$ that exactly $i$ of the children, $j$ females and $k$ males
in household $h$ become infected ($i=0,1,2$, $j=0,1$, $k=0,1$). The
expression is approximate because it explicitly depends upon the
numbers ultimately infected in the school and work places of individuals
in household $h$, and in the community.
Define $\bA_h$ as the random vector of numbers of children, females and
males ultimately infected in household $h$, so that $p_h(i,j,k) = P(\bA_h = (i,j,k))$.
To compute this, define $\ba_h$, the random vector denoting the numbers of children,
females and males infected from group or community infection in household $h$,
so that $\ba_h = (r,s,t)$ denotes the event that these numbers are $r$, $s$ and $t$, respectively.
Then
\begin{equation}
p_h(i,j,k) =\sum_{\begin{array}{c}0\le r\le i\\ 0\le s\le j\\0\le t\le
    k \end{array}
} P(\ba_h = (r,s,t))P(\bA_h = (i,j,k)| \ba_h = (r,s,t)).\label{lik-prob}
\end{equation}
The first factor on the right hand side of (\ref{lik-prob}) is
\[
P(\ba_h = (r,s,t)) =\binom{2}{r}\psi_c(h)^{2-r}\left(1-\psi_c(h)\right)^{r}\psi_f(h)^{1-s}\left(1-
  \psi_f(h)\right)^{s} \psi_m(h)^{1-t}\left(1-
  \psi_m(h)\right)^{t}.
\]
For the second factor on the right hand side of (\ref{lik-prob}), recall
that the within-household epidemic is homogeneous in the sense that
children and adults behave identically. Now if $i$ individuals are externally infected,
the probability of a final size of $k$ among the $s = 4-i$ initially susceptible
individuals is just $p_{4-i}^{(i)}(k)$, where $p_s^{(i)}(k)$ was
defined in Section 5.1.
Adjusting this probability to incorporate the extra information on which types
(children, female and male) were externally infected and which were eventually
infected amounts to multiplying by appropriate combinatorial factors. It follows that
\[
P(\bA_h = (i,j,k)| \ba_h = (r,s,t)) =
p_{4-(r+s+t)}^{(r+s+t)}(i-r+j-s+k-t)\frac{\binom{2-r}{i-r}
    \binom{1-s}{j-s}\binom{1-t}{k-t}}{\binom{4-(r+s+t)}{i+j+k-(r+s+t)}}.
\]
Note that $p_s^{(i)}(k)$ is a function of the parameter
$p_{\rm H}=1-e^{- \lambda_{\rm H} \mu }$.

If the final size data are summarized by $\{(i_h,j_h,k_h) \}_{h=1}^{n}$,
then the pseudolikelihood, which treats households as
independent, is
\[
L(  \{(i_h,j_h,k_h) \}_{h=1}^{n} ; \pi_{\rm C}, \pi_{\rm G}^{(1)}, \pi_{\rm G}^{(2)},
p_{\rm H}) =\prod_{h=1}^{n} p_h(i_h, j_h, k_h).
\]

For both complete and final size data, we proceed as for
Example 1, i.e.\ using similar MCMC methods to perform the
Bayesian analyses.

\subsection{Results}

As for Example 1, we consider two different datasets which illustrate
various key findings. In both cases the simulated outbreak is initiated by one index case who
is a child. For $j=0, \ldots, 4$, let $n_j$ be the number of households with final size $j$, so that $\sum_{j=0}^4 n_j = n = 500$.

{\bf Dataset 2.1} The first dataset was simulated using parameter
values $\lambda_{\rm H} = 0.3$, $\lambda_{\rm G}^{(1)} = 1.2$, $\lambda_{\rm G}^{(2)} = 0.6$,
and $\lambda_{\rm C} = 0.05$. It follows that $p_{\rm H} = 0.259$, $\pi_{\rm G}^{(1)} = 0.301$,
$\pi_{\rm G}^{(2)} = 0.549$, $\pi_{\rm C} = 0.951$, and $R_* = 2.08$. The parameter choices reflect high mixing in schools and moderate mixing at workplaces, households and the community at large.

In the simulated epidemic, being a typical one for the chosen parameter values, a total of 1475 individuals were infected, comprising
793 children, 333 females and 349 males, while $(n_0, n_1, n_2, n_3, n_4) = (25,40,79,147,209)$. Note that
the difference in numbers of males and females infected is purely random, since both genders experience
the same kind of mixing structure under the assumptions of the model.

{\bf Dataset 2.2} The second dataset was simulated using identical parameter values
to dataset 2.1 except that now $\lambda_{\rm C} = 0.005$, so that $\pi_{\rm C} = 0.995$ and
$R_* = 1.99$. As a consequence, the community mixing is now much lower.

The resulting simulated epidemic (also being typical) was far less severe, with 278 infected comprising 154 children, 68 females
and 56 males, and $(n_0, n_1, n_2, n_3, n_4) = (401,13,20,39,27)$.

\begin{table}
  \centering
  \caption{Posterior density summaries and ML estimates, Dataset 2.1}\label{ex2.1}
\vskip15pt
\begin{tabular}{l|c|c c c c |c c c c}
&True& \multicolumn{4}{|c}{Complete data} & \multicolumn{4}{|c}{Final size data}\\
&value& Mean & S. Dev. & Median & MLE & Mean & S. Dev. & Median & MLE\\
\hline
$p_H$ &0.259& 0.282 & 0.010 & 0.282 & 0.282 & 0.268 & 0.019 & 0.268 & 0.268\\
$\pi_{\rm G}^{(1)}$ &0.301& 0.289 & 0.016 & 0.289 & 0.289 & 0.288 & 0.031 & 0.285 & 0.268\\
$\pi_{\rm G}^{(2)}$ &0.544& 0.527 & 0.023 & 0.527 & 0.528 & 0.482 & 0.055 & 0.476 & 0.445\\
$\pi_{\rm C}$ &0.951& 0.951 & 0.001 & 0.952 & 0.953 & 0.933 & 0.062 & 0.952 & 1.000\\
$R_*$ &2.082& 2.280 & 0.092 & 2.278 & 2.268 & 2.339 & 0.107 & 2.337 & 2.340
\end{tabular}
\end{table}

\begin{table}
\centering
\caption{Posterior correlations, Dataset 2.1}\label{ex2.1cor}
\vskip15pt
\begin{tabular}{l|c|c}
& Complete data & Final size data\\
\hline
$\rho(p_H, \pi_{\rm G}^{(1)})$ & 0.063 & 0.30\\
$\rho(p_H, \pi_{\rm G}^{(2)})$ & 0.072 & 0.42\\
$\rho(p_H, \pi_{\rm C})$ & 0.0067 & 0.020\\
$\rho(\pi_{\rm G}^{(1)}, \pi_{\rm G}^{(2)})$ & 0.015 & 0.53\\
$\rho(\pi_{\rm G}^{(1)}, \pi_{\rm C})$ & -0.085 & -0.63\\
$\rho(\pi_{\rm G}^{(2)}, \pi_{\rm C})$ & -0.071 & -0.68\\
\end{tabular}

\end{table}

\begin{table}
  \centering
  \caption{Posterior density summaries and ML estimates, Dataset 2.2}\label{ex2.2}
\vskip15pt
\begin{tabular}{l|c|c c c c |c c c c}
&True& \multicolumn{4}{|c}{Complete data} & \multicolumn{4}{|c}{Final size data}\\
&value& Mean & S. Dev. & Median & MLE & Mean & S. Dev. & Median & MLE\\
\hline
$p_H$ &0.259& 0.247 & 0.021 & 0.246 & 0.246 & 0.237 & 0.032 & 0.238 & 0.235\\
$\pi_{\rm G}^{(1)}$ &0.301& 0.292 & 0.036 & 0.292 & 0.293 & 0.300 & 0.045 & 0.299 & 0.296\\
$\pi_{\rm G}^{(2)}$ &0.544& 0.538 & 0.049 & 0.524 & 0.541 & 0.526 & 0.076 & 0.523 & 0.519\\
$\pi_{\rm C}$ &0.995& 0.987 & 0.008 & 0.989 & 0.991 & 0.995 & 0.047 & 0.997 & 1.000\\
$R_*$ &1.992& 2.02 & 0.185 & 2.016 & 1.989 & 1.951 & 0.192 & 1.943 & 1.940
\end{tabular}
\end{table}

\begin{table}
\centering
\caption{Posterior correlations, Dataset 2.2}\label{ex2.2cor}
\vskip15pt
\begin{tabular}{l|c|c}
& Complete data & Final size data\\
\hline
$\rho(p_H, \pi_{\rm G}^{(1)})$ & 0.068 & 0.29\\
$\rho(p_H, \pi_{\rm G}^{(2)})$ & 0.050 & 0.53\\
$\rho(p_H, \pi_{\rm C})$ & 0.003 & 0.003\\
$\rho(\pi_{\rm G}^{(1)}, \pi_{\rm G}^{(2)})$ & 0.005 & 0.056\\
$\rho(\pi_{\rm G}^{(1)}, \pi_{\rm C})$ & -0.016 & -0.017\\
$\rho(\pi_{\rm G}^{(2)}, \pi_{\rm C})$ & -0.006 & -0.024\\
\end{tabular}

\end{table}

{\em General Remarks} Parameter estimates and correlations from analyses
of the two datasets are listed Tables \ref{ex2.1} - \ref{ex2.2cor}. Both
final size and complete data analyses yield broadly similar estimates, themselves
in line with what might be expected given the true values, all of which
suggests that the methods of inference are themselves effective.
Posterior standard deviations are lower for complete data, especially for the community
and workplace parameters, suggesting that temporal data have particular benefit
when it comes to estimating parameters associated with less common forms of
transmission. The gain in precision from having complete data is not as great
when it comes to estimation of $R_*$.

{\em Attribution of infection source} The results from Dataset 2.1 in Table
\ref{ex2.1} illustrate that temporal data can give a more accurate picture
of where transmission is occurring. Specifically, here the final size analysis
attributes less infection to within-household contact, and more to school and
community contact, than the complete data analysis.
The correlations in Table \ref{ex2.1cor} reinforce this finding, with all
but one of the final size analysis correlations being
of magnitude 0.3 or greater. As for Example 1, posterior correlations are
usually much smaller for complete data compared to final size data, illustrating
the utility of temporal data collection.

{\em How does outbreak size affect estimation?} Datasets 2.1 and 2.2 describe
very different epidemics: the former has attack rate (i.e.\ proportion infected)
74\%, the latter 14\%. Intuitively one would expect that more cases yield more
information about the model parameters. This is borne out by comparing the
posterior standard deviations in Tables \ref{ex2.1} and \ref{ex2.2}, the latter
generally being clearly larger than the former for both complete and final size
data. The one exception to this is the final size estimate of the community
parameter $\pi_{\rm C}$, which is estimated with higher precision given final
size data in the smaller outbreak rather than the larger. One possible
explanation for this arises from a more detailed consideration of Dataset 2.2,
and in particular the fact that the outbreak was confined to particular subsets
of the population. Specifically, only three of the ten schools had any cases,
and all of the adult cases arose in households with children at these schools.
Due to the structure of the population, individuals in households whose children
did not attend the infected schools had only community-level contact with
infected individuals. Consequently, 1400 individuals in the data are known to
have avoided infection due to community sources, leading to more precise estimation
of $\pi_{\rm C}$. Conversely, in Dataset 2.1 the outbreak is much larger and so it is
not possible to be so unambiguous about how individuals escaped infection in the final
size analysis. Note that although this argument could also be applied to the
complete data analysis, in that case there is already far more information about potential
sources of infection, which would explain why $\pi_{\rm C}$ is estimated with greater
precision in the larger outbreak.

{\em Data heterogeneity} As for Example 1, the posterior correlations in Tables \ref{ex2.1cor}
and \ref{ex2.2cor}, along with the arguments concerning $\pi_{\rm C}$ estimation above,
illustrate the fact that heterogeneity in the data can help estimation. The present
Example shows that this conclusion can still hold even when comparing outbreaks of very
different sizes.

\section{Discussion}
Although this paper is purely methodological, our aim has been to address
questions that are relevant for actual applied studies. In particular we have
tried to address broad questions of interest that are raised naturally when
one considers epidemic models which feature intermediate mixing levels.

We have derived methods for estimation based on final outcome and complete data,
and attempted to assess how different kinds of data enable estimation of key model
parameters. We found that temporal data can be of considerable benefit
compared to final outcome data, both in terms of the precision of estimates
and in terms of distinguishing different routes of infection. In particular it
was shown that having temporal information about the epidemic outbreak makes it
easier to distinguish community spread from that occurring within secondary group
structures. However, the extent to which the data are themselves homogeneous is
also an important factor in estimation. In particular, heterogeneity in the data
generally appears to improve estimation.

We have shown that estimation of the threshold parameter can be materially
changed when the intermediate level of mixing is ignored. This in turn implies
that estimation of vaccine coverage levels would also be affected by the
presence or otherwise of intermediate level mixing, since such levels are
usually determined by equating the threshold parameter to unity.

As with any model, some of our assumptions could be made more realistic.
For example, when considering adults and children it would be natural to allow these
two types of individual to have differing susceptibility or infectivity. In a similar
vein we have not taken account of any prior complete or partial immunity in the
population. Nevertheless it seems likely that the broad qualitative findings are unlikely
to be affected by such generalisations to the basic model considered here.

\section*{Acknowledgements}
This research has been partly funded by the Swedish Research Council, and by
UK Engineering and Physical Sciences Research Council grant EP/03234X/1.


\begin{thebibliography}{}

\bibitem[Andersen et~al., 1993]{abgk93}
Andersen, P.~K., Borgan, {\O}., Gill, R.~D., and Keiding, N. (1993).
\newblock {\em Statistical models based on counting processes}.
\newblock Springer Series in Statistics. Springer-Verlag, New York.

\bibitem[Anderson and May, 1991]{am91}
Anderson, R. and May, R. (1991).
\newblock {\em {Infectious diseases of humans: dynamics and control}}.
\newblock Oxford University Press, USA.

\bibitem[Andersson, 1999]{A99}
Andersson, H. (1999).
\newblock Epidemic models and social networks.
\newblock {\em Math. Sci.}, {\bf 24}, 128--147.

\bibitem[Andersson and Britton, 2000]{AB00}
Andersson, H. and Britton, T. (2000).
\newblock {\em Stochastic epidemic models and their statistical analysis},
  {\bf 151}, {\em Lecture Notes in Statistics}.
\newblock Springer-Verlag, New York.

\bibitem[\protect\citeauthoryear{Bailey}{Bailey}{1975}]{B75}
Bailey, N. T.~J. (1975).
\newblock {\em The Mathematical Theory of Infectious Diseases and its
  Applications\/} (Second ed.).
\newblock Hafner Press [Macmillan Publishing Co., Inc.]\ New York.

\bibitem[Ball and Lyne, 2001]{BL01}
Ball, F. and Lyne, O.~D. (2001).
\newblock Stochastic multitype {SIR} epidemics among a population partitioned
  into households.
\newblock {\em Adv. in Appl. Probab.}, {\bf 33}, 99--123.

\bibitem[Ball et~al., 1997]{BMS97}
Ball, F., Mollison, D., and Scalia-Tomba, G. (1997).
\newblock Epidemics with two levels of mixing.
\newblock {\em Ann. Appl. Probab.}, {\bf 7}, 46--89.

\bibitem[Britton and O'Neill, 2002]{BO02}
Britton, T. and O'Neill, P. D. (2002).
Bayesian inference for stochastic epidemics in populations with random social structure.
{\em Scand. J. Statist.} {\bf 29}, 375--390.

\bibitem[\protect\citeauthoryear{Cauchemez, Ferguson, Watchel, Tegnell, Saour, Duncan,
and Nicoll}{Cauchemez et~al.}{2009}]{CF09}
Cauchemez, S., Ferguson, N.M., Watchel, C., Tegnell, A., Saour G., Duncan B.
and Nicoll, A. (2009).
\newblock{Closure of schools during an influenxa pandemic.}
\newblock {\em Lancet Inf. Dis.} {\bf 9}, 473--481.

\bibitem[Cauchemez et~al., 2008]{CV08}
Cauchemez, S., Valleron, A., Bo{\"e}lle, P., Flahault, A., and Ferguson, N.
  (2008).
\newblock {Estimating the impact of school closure on influenza transmission
  from sentinel data}.
\newblock {\em Nature}, {\bf 452}, 750--754.

\bibitem[\protect\citeauthoryear{Clancy and O'Neill}{Clancy and
  O'Neill}{2008}]{CO08}
Clancy, D. and O'Neill, P. D. (2008).
\newblock {B}ayesian estimation of the basic reproduction number in stochastic epidemic models
\newblock {\em Bayesian Anal.\/}~{\bf 3\/}, 737--758.


\bibitem[Demiris and O'Neill, 2005]{DO05}
Demiris, N. and O'Neill, P.~D. (2005).
\newblock Bayesian inference for stochastic multitype epidemics in structured populations via random graphs
\newblock {\em J. Roy. Statist. Soc. Ser. B}, {\bf 67}, 731--746.

\bibitem[Diekmann and Heesterbeek, 2000]{DH00}
Diekmann, O. and Heesterbeek, J. (2000).
\newblock {\em {Mathematical epidemiology of infectious diseases: model
  building, analysis and interpretation}}.
\newblock Wiley,\ New York.

\bibitem[Ferguson et~al., 2005]{F05}
Ferguson, N., Cummings, D., Cauchemez, S., Fraser, C., Riley, S., Meeyai, A.,
  Iamsirithaworn, S., and Burke, D. (2005).
\newblock {Strategies for containing an emerging influenza pandemic in
  Southeast Asia}.
\newblock {\em Nature}, {\bf 437}, 209--214.

\bibitem[Halloran et~al., 2008]{HF08}
Halloran, M., Ferguson, N., Eubank, S., Longini, I., Cummings, D., Lewis, B.,
  Xu, S., Fraser, C., Vullikanti, A., Germann, T., et~al. (2008).
\newblock {Modeling targeted layered containment of an influenza pandemic in
  the United States}.
\newblock {\em Proc. Natl. Acad. Sci. USA}, {\bf 105}, 4639--4644 .

\bibitem[Keeling and Rohani, 2008]{KR07}
Keeling, M. and Rohani, P. (2008).
\newblock {\em Modeling infectious diseases in humans and animals}.
\newblock Princeton University Press, Princeton.

\bibitem[Kermack and McKendrick, 1927]{KM29}
Kermack, W.~O. and McKendrick, A.~G. (1927).
\newblock Contributions to the mathematical theory of epidemics, part i.
\newblock {\em Proc. Roy. Soc. London Ser. A}, {\bf 115}, 700--721.

\bibitem[Longini et~al., 2005]{L05}
Longini, I., Nizam, A., Xu, S., Ungchusak, K., Hanshaoworakul, W., Cummings,
  D., and Halloran, M. (2005).
\newblock {Containing pandemic influenza at the source}.
\newblock {\em Science}, {\bf 309}, 1083-1087.

\bibitem[Newman, 2003]{N03}
Newman, M. E.~J. (2003).
\newblock The structure and function of complex networks.
\newblock {\em SIAM Rev.}, {\bf 45}, 167--256 (electronic).

\bibitem[O'Neill, 2009]{ON09}
O'Neill, P. (2009).
\newblock Bayesian inference for stochastic multitype epidemics in structured
  populations using sample data.
\newblock {\em Biostatistics}, In Press.

\bibitem[O'Neill et~al., 2000]{OnBalBecEerMol00}
O'Neill, P.~D., Balding, D.~J., Becker, N.~G., Eerola, M., and Mollison, D.
  (2000).
\newblock Analyses of infectious disease data from household outbreaks by
  {M}arkov chain {M}onte {C}arlo methods.
\newblock {\em J. Roy. Statist. Soc. Ser. C}, {\bf 49}, 517--542.

\end{thebibliography}
\end{document}